\documentclass[12pt,preprint]{aastex}

\newcommand{\gsim}{\,\raisebox{0.2em}{$>$}\!\!\!\!\!\raisebox{-0.25em}{$\sim$}\,}
\newcommand{\lsim}{\,\raisebox{0.2em}{$<$}\!\!\!\!\!\raisebox{-0.25em}{$\sim$}\,}
\newcommand{\gr}{$\gamma$-ray \,}
\newcommand{\grs}{$\gamma$-rays \.}

\shorttitle{Contribution of Different SNRs into the Galactic Gamma-Ray Background}
\shortauthors{Berezhko \& V\"olk}

\begin{document}

\title{The Contribution of Different Supernova Populations to the
  Galactic Gamma-Ray Background}

\author{E.G. Berezhko}
\affil{Institute of Cosmophysical Research and Aeronomy, Lenin
Ave. 31, 677891 Yakutsk, Russia}
\email{berezhko@ikfia.ysn.ru}

\and

\author{H. J. V\"olk}
\affil{Max--Planck--Institut f\"ur Kernphysik, P.O. Box 103980,
D-69029 Heidelberg, Germany}
 \email{Heinrich.Voelk@mpi-hd.mpg.de}

\begin{abstract} 
The contribution of Source Cosmic Rays (SCRs), accelerated and still
confined in Supernova Remnants (SNRs), to the diffuse high energy \gr
emission above 1 GeV from the Galactic disk is studied. The \grs produced
by the SCRs have a much harder spectrum than those generated by the
Galactic Cosmic Rays (GCRs). Extending a previous paper, a simple SNR
population synthesis is considered and the Inverse Compton emission from
the SCR electrons is evaluated in greater detail. Then the combined
spectrum of \gr emission from the Galactic Supernova Remnant population is
calculated and this emission at low Galactic latitudes is compared with
the diffuse \gr emission observed by the EGRET and ground based
instruments.  The average contribution of SCRs is comparable to the GCR
contribution already at GeV energies, due to Supernovae of types II and Ib
exploding into the wind bubbles of quite massive progenitor stars, and
becomes dominant at \gr energies above $100$~GeV. At TeV energies the
dominant contribution is from SCRs in SNRs that expand into a uniform
interstellar medium. In fact, the sum of hadronic and Inverse Compton \grs
would exceed the limits given by the existing experimental data, unless
the confinement time $T_{SN}$, i.e. the time until which SNRs confine the
main fraction of accelerated SCRs, is as small as $T_{SN}\sim 10^4$~yr and
the typical magnetic field strength in SNRs as large as 30~$\mu$G.  Both
situations are however possible as a result of field amplification through
CR backreaction in the acceleration process. It is pointed out that
accurate measurements of the low-latitude diffuse Galactic \gr spectrum at
TeV-energies can serve as a unique consistency test for CR origin from the
Supernova Remnant population as a whole.
\end{abstract} 

\keywords{gamma-rays -- background radiation -- cosmic rays -- supernova
remnants} 

\section{Introduction} 

The diffuse Galactic \gr emission was measured with the Energetic Gamma
Ray Experiment Telescope (EGRET) on the Compton Gamma Ray Observatory up
to about 20 GeV. To first approximation it can be described by a suitable
model for the diffuse interstellar gas, the Galactic Cosmic Ray (GCR)
distribution, and the diffuse photon fields 
\citep{hunteretal97a,hunteretal97b}. 
The resulting GCR distribution in the Milky
Way essentially corresponds to that measured in situ in the local
neighborhood of the Solar System. This is also our starting point.

However above about 1~GeV the observed average diffuse \gr
intensity in the inner Galaxy, $300^{\circ}<l<60^{\circ}$,
$|b|\leq 10^{\circ}$, exceeds the model prediction significantly,
and this excess appears to increase monotonically with \gr energy.
Spatially, the diffuse \gr intensity is strongly increasing towards
the Galactic midplane.

In a previous paper \citet[hereafter referred to as Paper I]{bva00}, 
we have calculated the contribution of an unresolved distribution
of CR sources to the diffuse \gr flux in the disk, assuming the GCR
distribution within its large confinement volume to be the same as that
measured in situ. The CR sources contribute through the \gr emission of
the accelerated particles that are confined in their interior -- the
Source Cosmic Rays (SCRs). It was assumed that Supernova Remnants (SNRs)
are the dominant sources of the GCRs. On this premise it was found that
the SCRs give an important contribution to the diffuse high-energy
$\gamma$-ray emission at low Galactic latitudes. Since the CR energy
spectrum inside SNRs is much harder than on average in the Galaxy --- the
average spectrum being softened by rigidity-dependent escape from the
Galaxy in the diffusion/convection region above the disk --- the relative
SNR contribution increases with energy and was found to become in fact
dominant at \gr energies $\epsilon_{\gamma} \gsim 100$~GeV. This led to a
substantial increase of the "diffuse" TeV $\gamma$-ray emission from the
Galactic disk so as to constitute a significant observational background.  
This hard-spectrum background must also be taken into account in the
search for spatially extended Galactic CR sources above the GeV region.

Extending Paper I we take into account here that a significant fraction
of the SNR population, the type II and Ib Supernovae (SNe) with
progenitor masses in excess of $15M_{\odot}$, expand into a nonuniform
circumstellar medium strongly modified by the wind from the progenitor
star. In such SNRs the majority of CRs should be produced in a thin dense
shell of swept-up ISM. As a consequence of the high density of this shell
the expected $\pi^0$-decay $\gamma$-ray emission with energies
$\epsilon_{\gamma}\lsim 100$~GeV is shown to be an order of magnitude
larger than for an identical explosion into a uniform ISM. This gives a
natural explanation for the observed diffuse $\gamma$-ray excess for
energies $\gsim 1$~GeV. We also consider in detail the
contribution of the high-energy SCR electrons to the high energy \gr
background due to inverse Compton (IC) scattering on the background
photon field. Since the energy spectrum of the SCR electrons is strongly
influenced by synchrotron losses, the IC \gr emission depends on the
effective magnetic field strength in the SNR interior. This field may in
turn be amplified relative to the external field by the acceleration
process itself \citep{belll}. We take such a field amplification
into account as an average property of SNRs relative to the contribution
from the GCRs. As demonstrated below, this results in an order of
magnitude increase of the Galactic \gr emission at TeV-energies. In a
preliminary form these results have been presented in 
\citet{bv03}. 

We therefore conclude that a detailed measurement of the low-latitude
diffuse Galactic \gr spectrum within the energy interval from 1 to
$10^4$~GeV will constitute an important indirect test for the GCR
origin from the {\it population} of Galactic SNRs as a whole.

\section{Gamma-ray luminosity of SNRs expanding into a uniform circumstellar 
medium} 

The majority of the GCRs with charge number Z, at least up to kinetic
energies $\epsilon \sim 10^{15}Z$~eV, is presumably accelerated in SNRs.
Individual examples, where this is the case, are the objects SN 1006
\citep{bkv02,bkv03} 
and Cas A \citep{bpv03,bv04}. According to theory a significant part of the
hydrodynamic SN explosion energy $E_{SN}\approx 10^{51}$~erg is converted
into CRs already in the early Sedov phase of the evolution, $t\sim
10^3$~yr, as a result of diffusive shock acceleration
\citep{byk96,bv97,bvb00}. 
Later on the total CR energy
content $E_c$ and therefore the \gr production slowly varies with time. We
neglect this variation in the energetics in our considerations below,
assuming a time-independent SCR energy content $E_c=0.1E_{SN}$, consistent
with the GCR energy budget requirement.

The total number of active SNRs, $N_{SN}=\nu_{SN}T_{SN} \approx 300
~{\mathrm to}~ 3000$, is an increasing function of their assumed life time
$T_{SN} = 10^4~ {\mathrm to}~ 10^5$~yr, i.e. the time until which they can
confine the main fraction of the accelerated particles. Here $\nu_{SN}
\approx 1/30$~yr is the Galactic SN rate. Consequently we conclude that
the population of the oldest SNRs should dominate the total $\gamma$-ray
luminosity of the ensemble of Galactic SNRs. Approximately we shall
therefore consider only SNRs in the Sedov phase of their evolution in our
estimate of the total \gr luminosity.

Gamma-ray producing CRs in the Galaxy are then the sum of two basically
different populations. The first one consists of the ordinary GCRs
and presumably occupies a large Galactic residence volume
quasi-uniformly. This residence volume exceeds by far the volume
of the gas disk which harbors the CR sources 
\citep{ptuskinetal97,berezinetal90}. The
second CR population are the SCRs in the localized SNRs.

During the initial, active period of SNR evolution at times
$t\lsim T_{SN}$ when the SN shock is relatively strong, the volume
occupied by the accelerated CRs practically coincides with the
shock volume. During later evolutionary stages the shock becomes
weak and CRs begin to leave the SNR acceleration region. After
some period of time the escaping SCRs become very well mixed with
the ambient GCRs. Thus the controling factor is the shock
strength.

The $\gamma$-ray production by the GCRs is quite well studied
\citep{hunteretal97b,mori97}. Therefore we shall
concentrate on the relative contribution of the SCR population.

It was shown in Paper I that the total \gr spectrum measured from an
arbitrary Galactic disk volume is expected to be
\begin{equation} {dF_{\gamma}
\over d\epsilon_{\gamma}}= {dF_{\gamma}^{GCR}\over d\epsilon_{\gamma}}
[1.4+R(\epsilon_{\gamma})], 
\label{eq1}
\end{equation} 
where $dF_{\gamma}^{GCR}/d\epsilon_{\gamma}$ is the $\pi^0$-decay
\gr spectrum due to GCRs and the additional factor 0.4 is
introduced to approximately take into account the contribution of
GCR electron component to the diffuse \gr emission at GeV energies
\citep{hunteretal97b}.

For the ratio $R(\epsilon_{\gamma})=Q_{\gamma}^{SCR}/Q_{\gamma}^{GCR}$ of 
the $\gamma$-ray production rates due to SCRs and GCRs we have
\begin{equation} R(\epsilon_{\gamma})=0.07 \zeta
\left(\frac{T_p}{10^5~\mbox{yr}}\right)
\left(\frac{\epsilon_{\gamma}}{1~\mbox{GeV}}\right)^{0.6} (1+R_{ep}),
\label{eq2}
\end{equation} 
with $\zeta =N_g^{SCR}/N_g^{GCR}$, where $N_g^{GCR}=1 {\mathrm cm}^{-3}$
and $N_g^{SCR}$ are the gas number density in the Galactic disk and inside
SNRs, respectively, $R_{ep}=Q_{\gamma}^{IC}/Q_{\gamma}^{pp}$ is the ratio
of the IC to $\pi^0$-decay \gr production rates due to the SCRs, $T_p$ is
the proton SCR confinement time, and the difference between the spectral
indices of the GCRs and the SCRs is assumed to be 0.6.

In the case of SNRs expanding into a uniform ISM $T_p$, which is
equal to the confinement time $T_e$ of the SCR electron component
if synchrotron losses are not more restrictive for electrons (see
below), is given by the expression
\begin{equation}
T_p=\mbox{min}\{T_{SN},10^3(\epsilon/\epsilon_{max})^{-5}~\mbox{yr}\},
\label{eq3}
\end{equation} 
where $\epsilon_{max}=10^5 Z$~GeV is the maximal energy of CRs that can 
accelerated in
SNRs, and $\epsilon_{\gamma}=0.1\epsilon$ which is roughly valid for the
hadronic \gr production process. The decrease of the proton confinement
time $T_p$ with energy $\epsilon$ is due to the diminishing ability of the
SNR shock to produce high-energy CRs during the Sedov phase. In fact, the
highest CR energy $\epsilon$ and therefore the highest-energy $\gamma$-ray
production slowly decrease with time as $\epsilon \propto t^{-1/5}$ which
implies the escape of the highest energy CRs from the SNR 
\citep{ber96,byk96,bva00}.

Since it was shown in Paper I that electron SCRs give a \gr contribution
comparable with that of the nuclear SCRs, we consider the electron
emission here in greater detail. In the relevant energy range the \gr
luminosity due to the electron IC scattering on the background photons can
be described in the Thompson limit \citep{long81,berezinetal90}. 
Since the electron spectrum in a SNR depends significantly on the
SNR age $t$ due to synchrotron losses, we first determine the \gr
production rate $dQ_{\gamma}^{IC}(\epsilon_{\gamma})/dN_{SN}^e$ due to one
SNR of age $t$:
\begin{equation}
\frac{dQ_{\gamma}^{IC}(\epsilon_{\gamma})}{dN_{SN}^e} = \sigma_T c N_{ph}
\frac{d\epsilon_e}{d\epsilon_{\gamma}} 
\frac{dn^e_{SCR}(\epsilon_e)}{dN_{SN}^e} , 
\label{eq4}
\end{equation}
where 
\begin{equation} \epsilon_e=m_e
c^2\sqrt{3\epsilon_{\gamma}/(4\epsilon_{ph})} 
\label{eq5}
\end{equation} 
is the energy of electrons which produce an IC photon with mean
energy $\epsilon_{\gamma}$, $\sigma_T=6.65\times10^{-25}~{\rm
cm}^2$ is the Thomson cross section, and $\epsilon_{ph}$ and
$N_{ph}$ are the mean energy and number density of the background
photons, respectively. Finally
\begin{equation}
dn^e_{SCR}/dN_{SN}^e=N^e_{SCR}/V_g 
\label{eq6}
\end{equation} 
denotes the spatial electron SCR number density averaged over the
Galactic disk volume $V_g$, where $dN_{SN}^e=\nu_{SN}dt$ is
the number of SNRs of age between $t$ and $t+dt$ which contribute to
the IC emission at energy $\epsilon_{\gamma}$ from CR electron
sources.

The source electron spectrum at energies $\epsilon <\epsilon_l$,
where synchrotron losses are not important, can be represented in
the form
\begin{equation} 
N^e_{SCR}(t,\epsilon)=K_{ep}N_{SCR}(\epsilon),
\label{eq7}
\end{equation} 
where $K_{ep}\approx 10^{-2}$ is the electron to proton ratio and
$N_{SCR}(\epsilon) d\epsilon$ is the total number of SCR protons per SNR
in the energy interval d$\epsilon$. Finally
\begin{equation}
\epsilon_l=1.25 \left(\frac{10^5~\mbox{yr}}{t}\right)
\left(\frac{10~\mu\mbox{G}}{B}\right)^2~\mbox{TeV}
\label{eq8}
\end{equation}
is the low energy limit of that part of the electron spectrum
which is modified by the synchrotron losses, and $B$ is the
magnetic field strength inside the SNRs. For $\epsilon>\epsilon_l$
the electron spectrum is steeper due to synchrotron losses: 
\begin{equation}
N^e_{SCR}(t,\epsilon)=K_{ep}N_{SCR}(\epsilon)(\epsilon_l/\epsilon).
\label{eq9}
\end{equation} %

We also neglect the time dependence of the proton spectrum in
SNRs, since in the later Sedov phase which is most important
here, it is not significant, especially at high energies
\citep{byk96}. The proton spectrum is taken in a power
lower form $N_{SCR}\propto \epsilon^{-\gamma}$ with $\gamma=2.15$,
consistent with the requirements for the sources of the Galactic
CRs.

The maximum energy of SCR electrons is also restricted by their
synchrotron losses during their acceleration at the SNR shock:
\begin{equation}
\epsilon_{max}^e=24
\left(\frac{V_s}{10^3~\mbox{km/s}}\right)
\left(\frac{B}{10~\mu\mbox{G}}\right)^{-1/2}~\mbox{TeV},
\label{eq10}
\end{equation}
if $\epsilon^e_{max}<\epsilon_{max}$; here $V_s$ is the shock
speed. At the beginning of the
Sedov phase ($t\sim 10^3$~yr) the shock speed is about $V_s\approx
4\times 10^3$~km/s and subsequently decreases as $V_s\propto
t^{-3/5}$. Therefore the electron confinement time can be written
in the form
\begin{equation}
T_e=\mbox{min}\{T_p,10^3(\epsilon_e/\epsilon_{max}^e)^{-5/3}~\mbox{yr}\}.
\label{eq11}
\end{equation}
The total IC \gr production rate due to all existing SNRs can be
determined by the expression
\begin{equation}
Q_{\gamma}^{IC}=\int
_0^{T_e}\frac{dQ_{\gamma}^{IC}}{dN_{SN}^e}\nu_{SN}dt.
\label{eq12}
\end{equation}
First we normalize $Q_{\gamma}^{IC}$ to
the production rate of $\pi^0$-decay \grs from inelastic CR - gas
collisions, primarily p-p collisions, which may be written in the
form \citep{druryetal94}
\begin{equation}
Q^{pp}_{\gamma}(\epsilon)=Z_{\gamma}\sigma_{pp} c N_g n(\epsilon),
\label{eq13}
\end{equation} 
where $N_g$ is the local gas number density, $\sigma_{pp}$ is the
inelastic p-p cross-section, $Z_{\gamma}$ is the so-called
spectrum-weighted moment of the inelastic cross-section,
$n(\epsilon)=N_{SCR}(\epsilon)\nu_{SN}T_p/V_g$ is the spatial
number density of SCRs averaged over the Galactic volume $V_g$,
and $c$ is the speed of light.  Then, performing the integration
over all possible SNR ages, we find the ratio of luminosities: 
\begin{equation} R_{ep} = 73.6 K_{ep}
\left(\frac{N_{ph}}{N_g^{SCR}}\right) \left(\frac{4
\epsilon_{\gamma}\epsilon_{ph}} {3m_e^2c^4} \right)^{(\gamma-1)/2}
\label{eq14}
\end{equation} 
for \gr energies
$\epsilon_{\gamma}<\epsilon_{\gamma}^{*}$, and 
\begin{equation}
R_{ep} =73.6 K_{ep} \left(\frac{N_{ph}T_e}{N_g^{SCR}T_p}\right)
\left(\frac{4 \epsilon_{\gamma} \epsilon_{ph}} {3m_e^2c^4}
\right)^{(\gamma-1)/2}
\left(\frac{\epsilon_{\gamma}}{\epsilon_{\gamma}^*}\right)^{-1/2}
\left(1+\ln\sqrt{\frac{\epsilon_{\gamma}}{\epsilon_{\gamma}^*}}\right)
\label{eq15}  
\end{equation} 
for $\epsilon_{\gamma}>\epsilon_{\gamma}^{*}$, where
\begin{equation} \epsilon_{\gamma}^* \approx 8
\left(\frac{\epsilon_{ph}}{1~\mbox{eV}}\right)
\left(\frac{10^5~\mbox{yr}}{T_e}\right)^2 \left (\frac{B}{10~ \mu
\rm G}\right)^{-4}~\mbox{TeV} 
\label{eq16}
\end{equation} 
is the energy of \grs which are emitted by electrons with
synchrotron loss time equal to $T_e$. The value $Z_{\gamma}=0.113$
was used, which corresponds to $\gamma=2.15$ 
\citep{druryetal94}.

Due to the rather hard SCR electron spectrum the dominant
contribution to the IC radiation comes from collisions with cosmic
microwave background (CMB) photons which are characterized by
$N_{ph}=400$~cm$^{-3}$ and $\epsilon_{ph}=6.7\times10^{-4}$~eV.  
The far infrared radiation (FIR) field with $N_{ph} \approx
20$~cm$^{-3}$ and $\epsilon_{ph} \approx 0.01$~eV contributes
about 20\% for $\epsilon_{\gamma}\gsim 100$~GeV. In the case of
the FIR background the Klein-Nishina cross-section must be used
instead of the Thomson limit at TeV \gr energies. We take account
of the Klein-Nishina decrease of the cross-section at high
energies through multiplying $\sigma_T$ by the correction factor
\begin{equation}
f_{KN}=\exp (-3.5\sqrt{\epsilon_{\gamma} \epsilon_{ph}/m_e^2c^4}),
\label{eq17}
\end{equation}
which approximately describes the reduction effect \citep{blumg}. 
For $\epsilon_{\gamma}=10$~TeV,
$f_{KN}\approx 0.1$ for the FIR contribution.

Bremsstrahlung \grs play no role for the average \gr background
above GeV energies, if $K_{ep}\ll 0.1$ (Paper I).

In Fig.\ref{fig1} we present a calculated \gr spectrum based on the above
expressions (\ref{eq1})--(\ref{eq17}) with $T_{SN}=10^5$~yr, $B=10$~$\mu$G,
$\epsilon_{max}=10^5$~GeV, $K_{ep}=10^{-2}$ and $N_g^{SCR}=N_g^{GCR}$. A
SCR power law index $\gamma=2.15$ and a 10 percent efficiency of SCR
production in SNR, required to account for the GCRs, were used.

One can see that the SCR contribution becomes dominant for
$\epsilon_{\gamma}\gsim 100$~GeV where the expected \gr spectrum
becomes extremely hard. At TeV energies the predicted flux exceeds
the lowest HEGRA upper limit \citep{aha01} almost by a
factor of two. The SCR electron contribution increases with energy
$\epsilon_{\gamma}$ and exceeds the proton SCR contribution for
energies $\epsilon_{\gamma}\gsim 300$~GeV. At
$\epsilon_{\gamma}\sim 1$~TeV electron SCRs contribute about 60\%
of the total \gr flux. Therefore at these high energies the
expected \gr emissivity of SNRs is only weakly dependent on the
most relevant parameter, the maximum confinement time $T_{SN}$,
since according to expressions (\ref{eq2}) and (\ref{eq15})
we have for $\epsilon_{\gamma}\sim1$~TeV
\begin{equation}
dF_{\gamma}^{IC}/d\epsilon_{\gamma}\propto 1+0.3\ln(T_{SN}
/10^5~\mbox{yr}),
\label{eq18}
\end{equation}
taking into account that at this energy $T_p=T_e=T_{SN}$. Even if $T_{SN}$
is as short as $3\times 10^4$~yr, the expected SCR contribution at TeV
energies is therefore only 30\% lower and still exceeds the GCR
contribution by more than an order of magnitude (see Fig.\ref{fig1}). 
We emphasize
that in the latter case the SCR contribution is entirely due to the
electrons: they radiate about 85\% of the TeV \grs. This extreme case has
to be considered as the lowest value for the SCR contribution to the
Galactic \gr background radiation, since there are no other ways to
decrease it. This holds in particular also for the average gas number
density $N_g^{SCR}$ inside SNRs because the IC \gr production rate does
not depend on it. Such a short confinement time $T_{SN}\sim 10^4$~yr
\citep{pz} could be realized in a scenario where a
strong amplification of the magnetic field $B$ in SNRs is balanced by
nonlinear wave-wave interactions such that the wave magnetic power $P(l)$
in the spatial scale $l$ is dissipated in a (minimal) eddy turnover time
$\sim l/v(l)$ \citep{verm}. Here $v(l)=v_A [P(l)/B^2]^{1/2}$ is the
rms-value of the plasma mass velocity at scale $l$, and $v_A$ denotes the
Alfv$\acute{e}$n velocity.

We also point out that the restriction of the electron confinement time
$T_e$ due to their synchrotron losses becomes significant for \gr
energies $\epsilon_{\gamma}>1$~TeV and leads to the steepening of
the \gr spectrum (see Fig.\ref{fig1}).

According to Fig.\ref{fig1}, for energies $\epsilon_{\gamma}=0.4$
to 4~TeV the expected differential \gr energy spectrum from the
central part of the Galaxy has the form
\begin{equation}
dF_{\gamma}/d\epsilon_{\gamma}=A(\epsilon_{\gamma}/\mbox{1~TeV})^{-\alpha}
\label{eq19}
\end{equation}
with power law index $\alpha\approx 2.25$ and amplitude $A=(6~
{\mathrm to}~ 8)\times 10^{-9}$~(cm$^2$ s sr TeV)$^{-1}$, depending
on the confinement time $T_{SN}$.

As one can see from Fig.\ref{fig1}, the \gr flux expected for
$T_{SN}=10^5$~yr at $\epsilon_{\gamma}=1$~TeV exceeds the HEGRA upper
limit \citep{aha01} and the preliminary flux measured by the Milagro
detector \citep{fleysher}.  We emphasize here that we have multiplied the
Milagro flux by a factor of 3 since we present the expected \gr flux from
the region $|b|\leq2^{\circ}$ whereas the Milagro data correspond to
$|b|\leq5^{\circ}$. Even for $T_{SN}=3\times10^4$~yr the expected flux
exceeds the Milagro data.
 
The only physical parameter which strongly influences the IC \gr
production rate is the SNR magnetic field $B$: according to Eqs.
(\ref{eq15})--(\ref{eq16}) $F_{\gamma}^{IC}\propto B^{-2}$. A value of $B$ that
is significantly higher than $B=10$~$\mu$G can be attributed to field
amplification at the shock front due to the strong wave production by the
acceleration of CRs far into the nonlinear regime \citep{belll}.

In Fig.\ref{fig2} we present the same calculations as in
Fig.\ref{fig1} but with a mean SNR
magnetic field value $B=30$~$\mu$G. Compared with the previous case the IC
\gr emission is decreased by an order of magnitude. Therefore the
contribution of the hadronic SCRs to the \gr flux becomes much more
significant: at TeV-energies and for $T_{SN}=10^5$~yr the $\pi^0$-decay
\grs exceed the IC \grs by a factor of 2.5 , whereas for $T_{SN}=3\times
10^4$~yr the IC \grs still exceed the $\pi^0$-decay \grs by a factor of
1.3.  One can see that in this case the expected \gr flux is below the HEGRA
and Tibet upper limits already for a SCR confinement time
$T_{SN}=10^5$~yr. The expected \gr spectrum at energies
$\epsilon_{\gamma}=0.1~{\mathrm to}~10$~TeV is characterized by $\alpha
=2.4$ and $A=(2.4~ {\mathrm to}~ 4)\times 10^{-9}$~(cm$^2$ s sr TeV)$^{-1}$
, which also considerably exceeds the GCR contribution. Note that since
the maximum energy of SCRs obeys $\epsilon_{max}\propto B$, it is larger
by a factor of three compared with the previous case. Taking into account
the uncertainty of the Milagro flux one can conclude that this measurement
does not contradict our prediction for the case of the amplified magnetic
field. The lower value of the confinement time $T_{SN}\sim 10^4$~yr might
be preferable in this sense.

\section{Wind SNRs of type II and Ib}

The above consideration corresponds to the assumption that all existing
Galactic SNRs expand into a uniform circumstellar medium with a density of
$N_g^{GCR}=1$~cm$^{-3}$.  However, for stellar masses in excess of about
15 $M_{\odot}$ the progenitors of type II and Ib SNe can strongly modify
their environment through their ionizing radiation and very energetic
winds \citep{weaver,chevl}.
Disregarding large-scale turbulent mixing processes in the sequel, the
stellar wind from such a star creates a low-density bubble beyond the
termination shock out to a radius $R_{sh}$, bounded by a dense swept-up
shell of interstellar material. The radius of the wind bubble can be
several tens of parsecs. The very hot dilute bubble contains a small
amount of mass, typically a few solar masses or even less, whereas the
thin swept-up shell ultimately contains thousands of solar masses. This
shell will have a gradual inner boundary where
localized mixing of bubble and shell material proceeds. Since the typical
SN ejecta mass is about $M_{ej}\approx 10M_{\odot}$ for massive
progenitors, only a relatively small fraction of the SN explosion energy
$E_{SN}$ is transformed into internal gas and CR energy during the SN
shock propagation through the bubble and the main part of SN energy is
deposited in the shell.

There are no physical reasons for which the
CR production process in the shell matter is substantially different from
that in the uniform ISM case, apart from the different distribution of ion
injection across the SNR shock surface due to the magnetic field geometry
\citep{vbk03}. 
The major difference which plays a role for the
$\pi^0$-decay \grs is that CRs accelerated during the SN shock propagation
through the shell are confined in a medium which is more than a factor of
ten denser than the average ISM in the Galactic disk, of density
$N_g^{GCR}$. This may lead to an appreciable increase of the overall \gr
production.

In addition, the SNR shock evolves differently than in
the case of a uniform circumstellar medium. The entire active
period of SNR evolution, when the SN shock is strong enough and
effectively produces CRs, takes place within the shell of
thickness $L$ which is much smaller than its radius, $L\ll
R_{sh}$. This implies that during the entire active evolution the
SNR shock size $R_s$ remains nearly constant. The time dependence
of the shock speed $V_s$ which is very important for the SCR
confinement in SNRs can then be estimated as follows.

The majority of the progenitor stars with an intense wind are
main-sequence stars with initial masses $M_i>15M_{\odot}$ 
\citep{abb}. In the mean, during their evolution in the surrounding uniform ISM
of gas number density $\rho_0=m_pN_g^{GCR}$, they create a bubble of size
\citep{weaver,chevl}
\begin{equation}
R_{sh}=0.76( 0.5 \dot{M}V_w^2 t_w^3/\rho_0)^{1/5},
\label{eq20}
\end{equation}
where $\dot{M}$ is the mass-loss rate of the presupernova star, $V_w$ is the 
wind speed, and $t_w$ is the duration of the wind period. 

In order to determine the SN shock dynamics inside the shell we model the
gas number density distribution in the bubble and in the shell in the form
\begin{equation}
N_g=N_b+(r/R_{sh})^{3(\sigma -1)}N_{sh},
\label{eq21}
\end{equation}
where $N_{sh}=\sigma N_g^{GCR}$ is the peak number density in the shell,
$N_b$ is the gas number density inside the bubble, typically very
small compared with the shell density, and $\sigma =N_{sh}/N_0$ is
the shell compression ratio.

The mass of the bubble 
\begin{equation}
M_b=(4\pi R_{sh}^3/3)m_pN_b
\label{eq22}
\end{equation}
is about $3~ {\mathrm to}~ 5 M_{\odot}$, whereas the shell mass
\begin{equation}
M_{sh}=4\pi N_{sh}m_p\int_0^{R_{sh}}dr r^2(r/R_{sh})^{3(\sigma -1)}=
(4\pi R_{sh}^3/3)N_g^{GCR}m_p
\label{eq23}
\end{equation}
is a few thousands of solar masses. Therefore, during SNR shock
propagation through the bubble, only a small fraction of its energy is
given to gas of stellar origin. The main part of the explosion energy is
deposited in the shell.

As in the case of a uniform ISM the evolution of the SN shock can
be represented as a sequence of two stages. 

The first stage is the free expansion phase which continues up to the
time, when the shock sweeps up a mass equal to the ejecta mass $M_{ej}$.
The second phase is analogous to the Sedov phase in the uniform ISM. The
only significant difference compared with the uniform ISM case is that at
the end of the free expansion phase, when a large part of the SN energy
has been given to the swept up gas, the shock volume is much larger.

When the swept-up mass becomes large compared to the sum of bubble
mass and $M_{ej}$, the SN shock propagation in the shell medium
with density $N_g\propto r^{3(\sigma-1)}$ can be shown to approach
the following
self-similar adiabatic solution
\begin{equation}
R_s^3V_s^2N_g(R_s)=\mbox{const},
\label{eq24}
\end{equation}
which corresponds to an expansion law $R_s=R_0(t/t_0)^{\nu}$ with
$\nu=2/(3\sigma +2)$. The shock decelerates much more rapidly in
this case, $V_s\propto t^{-3\sigma/(3\sigma+2)}$, than in the case
of a uniform ISM, where $V_s\propto t^{-3/5}$. Due to the fact that
$3\sigma\gg 1$ we have approximately $V_s\propto t^{-1}$. 
During this period of time the shock size and the amount of
the swept-up mass are connected by the relation
\begin{equation}
R_s=R_{sh}(M/M_{sh})^{1/3\sigma}.
\label{eq25}
\end{equation} 

Substituting in this expression the SN ejecta mass $M_{ej}$ we get the
value of the SN shock size which corresponds to the end of free
expansion phase
\begin{equation}
R_s(t_0)=R_0=R_{sh}(M_{sh}/M_{ej})^{-1/3\sigma}.
\label{eq26}
\end{equation}
At
this epoch the shock speed is roughly the mean ejecta speed
\begin{equation}
V_s\approx V_0=\sqrt{2E_{SN}/M_{ej}}.
\label{eq27}
\end{equation}
Therefore the corresponding time scale is $t_0=R_0/V_0$. 

As it follows from the expansion law and from eq.(\ref{eq24}), the SNR
shock reaches the edge of the shell $R_s(t_f)=R_{sh}$
at time 
\begin{equation}
t_f=t_0(M_{sh}/M_{ej})^{(3\sigma+2)/6\sigma},
\label{eq28}
\end{equation}
which for $\sigma\gg 1$ gives
\begin{equation}
t_f \approx t_0\sqrt{M_{sh}/M_{ej}}.
\label{eq29}
\end{equation}
At about this stage the SNR shock has lost much of its speed and will 
come into pressure equilibrium with its environment, i.e. $t_f \approx 
T_{SN}$.

For the typical values $\dot{M}=6\times 10^{-9}M_{\odot}$/yr,
$V_w=2500$~km/s, $t_w=10^7$~yr, which correspond to a progenitor of mass
$M_i=15M_{\odot}$ \citep{abb,chevl}  and
$N_g^{GCR}=1$~cm$^{-3}$, we have $R_{sh}\approx 30$~pc and $M_{sh}=3\times
10^3M_{\odot}$. For $E_{SN}=2\times 10^{51}$~erg and $M_{ej}=5M_{\odot}$
this gives $t_0=5\times 10^3$~yr and $t_f\approx 10^5$~yr. The ambient
interstellar gas may also be photoionized which implies a large enough
external pressure to stop bubble expansion at a lower radius
\citep{chevl}. 
Therefore we also consider below the case
$R_{sh}=17$~pc which corresponds to $t_f=3\times 10^4$~yr.

According to theory, the expanding SNR shock produces a power law
CR spectrum up to a maximum energy \citep{ber96,byk96,bv97}
\begin{equation}
\epsilon_m\propto R_sV_s,
\label{eq30}
\end{equation}
which is determined by the radius $R_s$ and speed $V_s$ of the
shock. The CRs with the highest energy $\epsilon_{max}$ are
produced at the very beginning of the Sedov phase $t\sim t_0$ when
the product $R_sV_s$ has its maximum. Subsequently, the product
$R_sV_s$ decreases with time approximately as $t^{-1}$ and the SNR
shock produces CRs with progressively lower cutoff energy
$\epsilon_m(t)<\epsilon_{max}=\epsilon_m(t_0)$. During that phase
those CRs that were previously produced with energies
$\epsilon_m<\epsilon<\epsilon_{max}$ now leave the remnant without
a significant influence of the SNR shock. Therefore the CR
confinement time has to be taken in the form
\begin{equation}
T_p=min\{1,(t_0/t_f)(\epsilon_{max}/\epsilon)\}T_{SN},
\label{eq31}
\end{equation} 
where according to the above consideration $T_{SN}=t_f$. 
This relation is analogous to eq.(\ref{eq3}) in the case of a
uniform ISM and leads to an increase of the spectral index of the 
\gr emission by one unit.

In order to estimate the highest energy of accelerated CRs in this case we
use the relation $\epsilon_{max}\propto R_0 V_0 B$. Compared with the case
of a uniform ISM, where $R_0=4$~pc for $N_g^{GCR}=1$~cm$^{-3}$, the value
of $R_0$ is about ten times larger but $V_0$ is the same. The magnetic
field $B(t_0)$ on the other hand is much lower than in the case of a
SN~Ia.  Indeed, as follows from eqs. (\ref{eq25}) and (\ref{eq20}), the gas number density
at the beginning of the Sedov phase is as small as
$N_g(t_0)=N_{sh}M_{ej}/M_{sh}\approx N_g^{GCR}/300$. The magnetic field in
the region of the shell is presumably adiabatically
compressed interstellar field $B_0$. If we approximate the directions of
$B_0$ as being isotropic, then its magnitude scales with the gas density
like $B\propto N_g^{2/3}$, e.g. \citet{chev74}. This gives a field
strength $B(t_0)\approx 0.02B_0$. Taking into account all the factors
considered, we conclude that the maximum CR energy in the case of wind-SNe
with dense shells is roughly a factor of ten smaller than in the
case of SNe expanding into a uniform ISM. Assuming that the field
amplification is proportional to $\sqrt{N_g}$, cf. \citet{belll},
this amplification does not play a role here since
$N_g(t_0)$ is so low.

In contrast to our previous study \citep{bvb00}, where we
considered the CR acceleration by SN shocks expanding into a {\it 
modified}
bubble, we consider here the opposite extreme of bubble structure. It is
assumed that the magnetic field suppresses the mass and heat transport
between the dense shell and the hot bubble \citep{chevl}.
Therefore the bubble mass is so small that a significant number of CRs is
produced by the SN shock only when it enters the dense shell. Even though
it is not clear at the moment which of these two concepts is physically
more correct, we consider here
the possibility of an unmodified bubble, because of its great importance
for \gr production.

We note that due to the energy dependence of the confinement time
$T_p(\epsilon)$ the SCR contribution to the \gr flux undergoes a spectral
break at a \gr energy which corresponds to the SCR energy
\begin{equation}
\epsilon_{break}=(t_0/t_f)\epsilon_{max}.
\label{eq32}
\end{equation}
Protons with energies $\epsilon\le \epsilon_{break}$ survive up to the
final active SNR epoch $t=t_f$ and therefore provide the maximum
contribution to the \gr SNR emissivity at
$\epsilon_{\gamma}^{break}=0.1\epsilon_{break}$, whereas for
$\epsilon_{\gamma}>\epsilon_{\gamma}^{break}$ the \gr emissivity decreases
more steeply, to go to zero at $\epsilon_{\gamma}=0.1\epsilon_{max}$.

The gas number density $N_g^{SCR}$ in expression (\ref{eq2}) is a function of SNR
age $t$ for wind-SNe, because the gas density $N_g (R_s)$ seen by the SN
shock changes during the evolution of the SNR. It then follows from eq.  
(\ref{eq20})  that the gas density behind the SN shock, where most of the SCRs are
located, can be represented as $N_2(t)=\sigma_{SN} N_g(R_s(t))=\sigma_{SN}
(t/t_f)^2N_{sh}$, where $\sigma_{SN}$
is the SN shock compression ratio. Therefore the gas density seen by SCRs
of energy $\epsilon$ is
\begin{equation} 
N_g^{SCR}=\sigma_{SN}(T_p/T_{SN})^2 N_{sh}.
\end{equation} 
Since $T_p\propto \epsilon_{\gamma}^{-1}$ for
$\epsilon_{\gamma}>\epsilon_{\gamma}^{break}$, the \gr spectrum produced
in the shell is very steep at these high energies, and in 
eq. (\ref{eq2}) we have
$R(\epsilon_{\gamma})\propto \epsilon_{\gamma}^{-2.4}$.

We assume that the progenitors of type II SNe are stars more massive than
$M_i>8M_{\odot}$, and that only those with $M_i>15M_{\odot}$ have strong
winds which produce extended bubbles, e.g. \citet{abb}. This is also
true for the progenitor population of type Ib SNe. According to 
\citet{gm} the initial mass function has the form $dN/dM_i\propto
M_i^{-\alpha}$ with $\alpha=1.6, 2.4, 3.24, 3.62$ for the mass intervals
$M_i<1 M_{\odot}$, $1M_{\odot}<M_i<10 M_{\odot}$, $10M_{\odot}<M_i<50
M_{\odot}$, $50M_{\odot}<M_i$, respectively. Therefore the ratio of the
number of stars whose initial mass $M_i$ exceeds 15~$M_{\odot}$ to those
with $M_i> 8 M_{\odot}$ is 0.23. Since about 85\% of all SN explosions in
the Galaxy are type II and type Ib SNe \citep{tamm}, and since in
addition the explosion energies $E_{SN}$ of type~II and type~Ib SNe with
$M_i>15M_{\odot}$ are about two times larger than on average 
\citep{chev,hamuy}, we conclude that roughly a fraction $\delta=0.3$ of
the total Galactic SN energy release is from wind-SNe with progenitors of
masses $M_i>15M_{\odot}$ which produce extended bubbles.

Therefore the expected \gr SNR luminosity should be weighted
between the SNR populations expanding into a uniform ISM and into bubbles 
as follows:
\begin{equation}
R=(1-\delta)R_I+\delta R_{II},
\label{eq34}
\end{equation}
where $R_I$ corresponds to the ratio of \gr production rates due to SCRs
and GCRs if all SNe were exploding into a uniform ISM, and $R_{II}$ is the
corresponding ratio for explosions into bubbles. 

Assuming that wind-SNe have rarefied bubbles with dense
shells the spectrum of \grs calculated with the above ratio $R$,
$\sigma=10$, $\sigma_{SN}=5$ and $\delta =0.3$ is presented in
Figs.\ref{fig1} and
\ref{fig2} by the full lines, which correspond to two different assumptions about
the value of the SCR confinement time $T_{SN}$. It is seen that the
wind-SNe dominate at energies $\epsilon_{\gamma}<100$~GeV and fit the
EGRET data for $T_{SN}=3\times 10^4$~yr fairly well, whereas for the
larger confinement times $T_{SN}=10^5$~yr their contribution becomes so
large, that their calculated flux exceeds the EGRET data by about a factor
of two.

We emphasize that the contribution from the population of wind-SNe is
mainly due to the nuclear SCR component. The break in the \gr spectra
presented in Fig.\ref{fig1}, \ref{fig2} 
is at $\epsilon_{\gamma}^{break}=30$~GeV for
$T_{SN}=10^5$~yr and at $\epsilon_{\gamma}^{break}=100$~GeV for
$T_{SN}=3\times 10^4$~yr. For larger energies the proton confinement time
and the gas density become lower, $T_p\propto \epsilon^{-1}$,
$N_g^{SCR}\propto \epsilon^{-2}$, which leads to a corresponding decrease
of the wind-SN contribution to the \gr spectrum. This contribution becomes
insignificant for $\epsilon_{\gamma}\sim 1$~TeV.

\section{Discussion}

Our above consideration is based upon a picture in which the population of
SNRs, considered here as the main GCR sources, are distributed across the
Galactic disk similar to the Interstellar gas. In this case the calculated
ratio $R(\epsilon_{\gamma})$ of the SCR to the GCR contribution does not
depend upon the observing direction in the disk. Since the number of these
sources $N_{SN}=\nu_{SN}T_{SN}$ is very limited, their mean number within
the field of view $\sim {\mathrm few}$~ degrees of a stereoscopic system
of imaging atmospheric Cherenkov telescopes for \gr energies $\gsim
100$~GeV is so low, that one should expect large fluctuations of the
actual value of $R(\epsilon_{\gamma})$, especially for lines-of-sight
other than those directed towards the inner Galaxy.

The actual SNR distribution within the Galactic disk most probably is not
uniform. According to \citet{case} the SN explosion rate as a function of
Galactocentric radius $r$ has a peak at $r\approx 5$~kpc and drops
exponentially with a scale length of $\sim 7$~kpc. Within $5<r< 20$~kpc
this agrees fairly well with the radial distributions of {\it Supernovae}
in a sample of 36 {\it external galaxies} of Hubble type Sb--Sbc,
statistically considered equivalent to our Galaxy \citep{drag}.  From our
results, for \gr energies larger than 10 GeV, the ratio $R$, corrected for
the nonuniform SNR distribution, is then given by the ratio $R_c=\eta(r)
R$, where the parameter $\eta(r)$ represents the ratio of the actual SNR
number along the given line-of-sight to that SNR number which would
corresponds to their uniform distribution. For lower \gr energies, $100<
\epsilon_{\gamma}< 10^4$~MeV, where the truly diffuse emission from the
GCRs dominates, it is known that the \gr emissivity gradient is
significantly shallower than the gradient of $\eta(r)$, as shown by the
EGRET instrument on board of the CGRO satellite \citep{strm} in basic
agreement with the results from the COS-B satellite \citep{stretal}. This
much shallower, indeed truly diffuse Galactocentric \gr gradient can be
understood as a nonlinear propagation effect from the disk into the
Galactic Wind where the latter is driven by the GCRs themselves
\citep{breit}. In contrast, the radial galactocentric variation of the
direct radiation from the sources is not smoothed by any propagation
effect. Therefore its amplitude should spatially vary like the
distribution of the sources, integrated along the line-of-sight, and this
variation should become directly measurable in the $> 100$~GeV range.
Given several such line-of-sight integrals, one can infer the radial
galactocentric \gr emissivity gradient. The excellent angular resolution
of atmospheric Cherenkov telescopes and the high sensitivity of
instruments like the CANGAROO III and H.E.S.S. arrays in the Southern
Hemisphere, which at TeV energies have an order of magnitude higher
sensitivity compared with previous generation instruments like HEGRA
\citep{aha01}, should allow this measurement in particular in the inner
Galaxy.

The assumed average value $B=30~\mu$G of the magnetic field inside SNRs
is considerably smaller than the value $120~\mu$G inferred for e.g.
SN~1006 in the very early Sedov phase \citep{bkv02,bkv03}. 
However the amplification must be a monotonically increasing
function of the shock strength. In particular, \citet{belll} argue
for a dependence $B \propto V_s$. If so, then the effective amplified
magnetic field in SNRs varies from $B\sim 100$~$\mu$G at the beginning of
the Sedov phase to the typical interstellar value $B\approx 10$~$\mu$G at
the end of its active period $t=T_{SN}$. Since the value of the magnetic
field in SNRs has a dominant influence on the IC \gr spectrum, and since
all evolutionary phases of a SNR contribute to this spectrum in a roughly
equal way, the adopted value $B=30$~$\mu$G appears to be quite realistic,
averaging over the entire active period.

One should also note that the SNR magnetic field influences the maximum
energy of SCRs. Since the most energetic SCRs are created at the beginning
of the Sedov phase, their actual maximum energy $\epsilon_{max}\propto
B(t_0)$ is expected to be larger than considered here, because the
amplified magnetic field $B(t_0)$ in this phase is by a factor of several
larger than on average. Compared with the Fig.\ref{fig2} this effect will produce
more a extended and smooth high energy tail of the \gr spectrum up to the
energy $\epsilon_{\gamma}\approx 10^5$~GeV.

The actual process of SN shock interaction with the dense shell could be
much more complicated compared with the ideal picture considered here. If
due to some physical factors (inhomogeneities in the surrounding ISM,
instabilities ...) the shell is {\it always} strongly distorted, then one
would expect that the SN shock will penetrate through the shell more
rapidly. This will lead to a decrease of the propagation time $t_f$
and/or to the decrease of the effective gas number density $N_g^{SCR}$,
that in turn will decrease the amount of \grs produced in such types of
SNRs. In other words, the calculated \gr spectrum at
$\epsilon_{\gamma}\lsim 100$~GeV (see Fig.\ref{fig1}, \ref{fig2}), 
which is mainly due to
the wind SNe, has to be considered as an upper limit, which corresponds
to the case of stable unmodified bubbles.

\section{Summary}

Our considerations demonstrate that the SCRs inevitably make a strong
contribution to the ``diffuse'' \gr flux from the Galactic disk at all
energies above a few GeV, if the population of SNRs is the main
source of the GCRs. According to our estimates, the SCR contribution
dominates at energies greater than 100~GeV
due to its substantially harder spectrum.

There are two physical parameters which influence the expected
\gr emission from SNRs significantly: the CR confinement time and the mean
magnetic field strength $B$ inside the SNRs. For a conventional value
$B=10$~$\mu$G the expected \gr flux from SNRs exceeds the HEGRA upper
limit considerably if the SCR confinement time is as large as
$T_{SN}=10^5$~yr. This contradiction can be resolved either if we suggest
an appreciably higher postshock magnetic field $B\gsim 30$~$\mu$G or if
the SCR confinement time is as small as $T_{SN}\sim 10^4$~yr, or a
combination of both effects. These possibilities can be attributed to
field amplification by the SCRs themselves. In fact, nonlinear field
amplification may also lead to a substantial decrease of the SCR
confinement time: according to \citet{pz} maximal
turbulent Alfv$\acute{e}$n wave damping with its corresponding increase of
CR mobility could make the SCR confinement time as small as $T_{SN}\sim
10^4$~yr. Under these circumstances the most realistic \gr background
spectrum is represented by the thick solid line in Fig.\ref{fig2}. 
One can see that
even in the case of $B= 30$~$\mu$G and $T_{SN}\sim 3\times 10^4$~yr the
SNR contribution at TeV energies, with roughly numbers of IC and $\pi^0$-decay
\grs, still exceeds the GCR contribution almost by an order of
magnitude. Note that the preliminary \gr flux at
$\epsilon_{\gamma}=1$~TeV, measured by the Milagro detector 
\citep{fleysher}, 
confirms our earlier prediction \citep{bv03}.

\begin{figure}
\epsscale{0.80}
\plotone{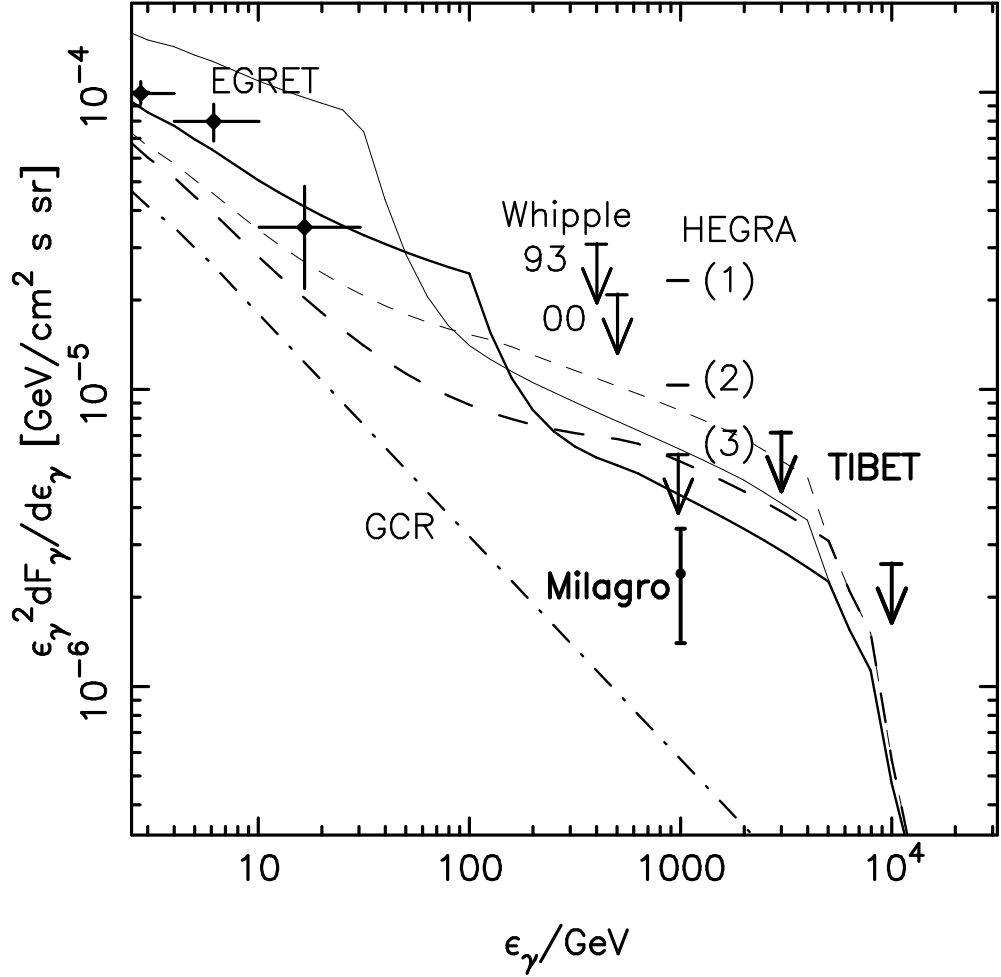}
\caption{
The average diffuse \gr spectrum of the low latitude inner Galaxy
($38^{\circ}<l<43^{\circ}$, $|b|\leq2^{\circ}$). The thick and the
thin dashed lines represent the case of a uniform ISM with the two
different SNR life times $T_{SN}=10^5$~yr and $T_{SN}=3\times
10^4$~yr, respectively; the full lines correspond to the case when
30\% of the SN energy release is due to SNRs which expand into the
wind bubble of a massive progenitor star. EGRET data
\citep{hunteretal97b}, 
preliminary Milagro data at $\epsilon_{\gamma}=1$~TeV
\citep{fleysher}, 
the Wipple upper limits \citep{reyn,leboh}, the HEGRA upper limits \citep{aha01}, and
the Tibet upper limits \citep{tibet} at
$\epsilon_{\gamma}=3$, 10~TeV are shown. The Milagro data have been
multiplied by a factor of 3 (see text).
\label{fig1} }
\end{figure}

\begin{figure}
\plotone{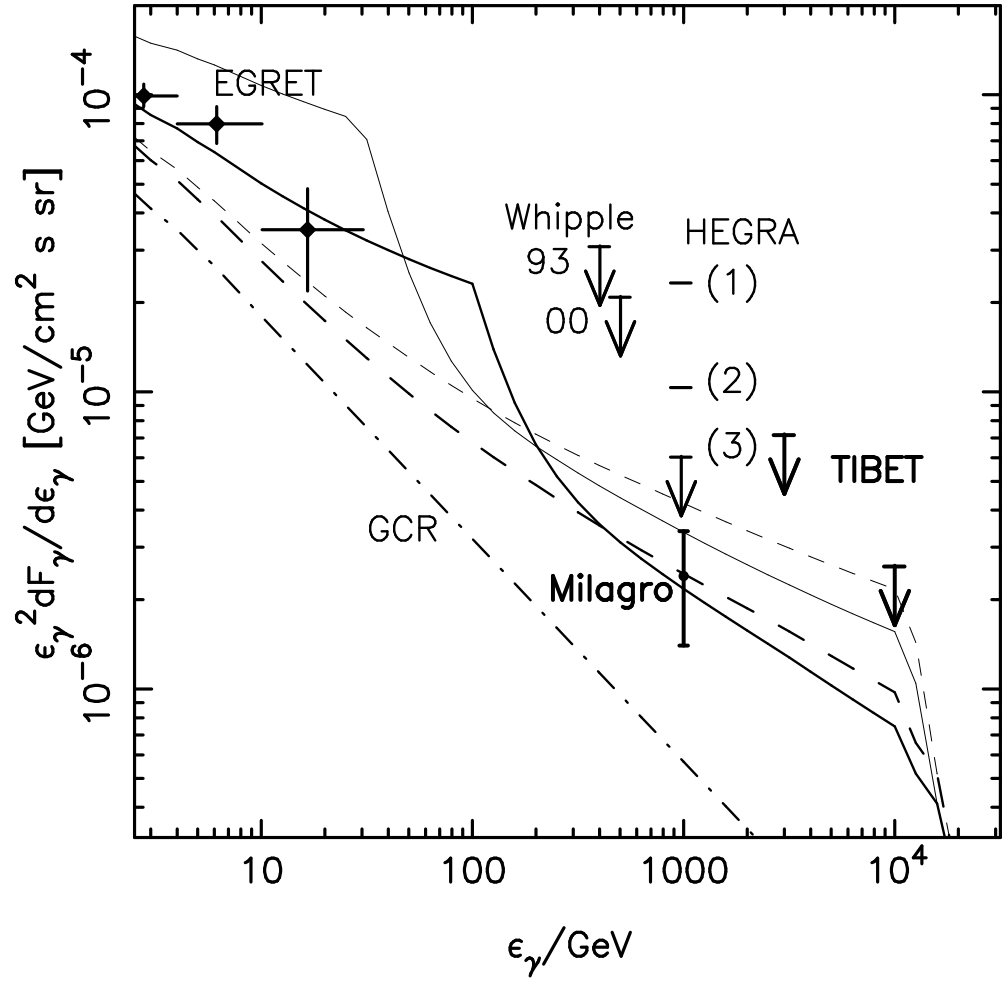}
\caption{
The same as in Fig.1 but with the postshock SNR 
magnetic field value $B=30$~$\mu$G.
\label{fig2} }
\end{figure}

At lower energies $\epsilon_{\gamma}\lsim 100$~GeV the \gr emission from
SNRs is dominated by the wind-SNe with initial progenitor mass
$M_i>15M_{\odot}$ which expand into the bubble created by the progenitor's
wind, under the assumption that the bubble is not modified by global
heat and mass transport. The main fraction of the CRs is produced in this
case when the SN shock propagates through the thin dense shell at the edge
of the bubble. Due to the significantly higher mass density of the shell
material compared with the mean gas density in the Galactic disk
$\pi^0$-decay \grs dominate at \gr energies $\epsilon_{\gamma}\lsim
100$~GeV despite the fact that only about 20\% of all SNRs belong to this
class of objects. As shown in Figs.\ref{fig1} and \ref{fig2}, 
the discrepancy between the
observed ``diffuse'' intensity and standard model predictions at energies
above a few GeV can be attributed to the SCR contribution. This
requires unmodified bubbles for the wind-SNe. Since we
cannot prove this assumption, the corresponding explanation of the lower
energy \gr excess is not a definitive conclusion, but rather a
plausible suggestion which must await more detailed studies of wind bubble
morphologies.

We conclude that a detailed measurement of the low-latitude ``diffuse''
Galactic \gr {\it spectrum} within the energy interval from 1 to
$10^4$~GeV 
will allow a strong consistency check for the predominant origin of the
Galactic Cosmic Rays from the Galactic population of SNRs as a whole. In
the energy range above a few 100 GeV our predictions are quite robust
resulting in a hard spectrum whose Galactocentric variation should
correspond to that of the observed SNR distribution.
If the preliminary Milagro data, which agree with our prediction, 
are confirmed by more precise measurements, then such measurements
will give an indirect confirmation of SNRs as the main sources
of GCRs. The measurements will also give
a new tool to study the spatial distribution 
of SNRs in the Galaxy.

\acknowledgments

This work has been supported in part by
the Russian Foundation of Basic Research, grants 03-02-16524 and by LSS
grant 422.2003.2. We thank G. P\"uhlhofer for a discussion on the
observations of diffuse $\gamma$-rays. EGB acknowledges the hospitality of the
Max-Plank-Institut f\"ur Kernphysik where part of this work was carried
out.

\clearpage

%fig1

\clearpage
%\newpage

%fig2

\end{document}